\documentclass[conference]{IEEEtran}
\usepackage{myStyleIEEE}
\usepackage{ushort}
\usepackage{threeparttable}
\usepackage{makecell}
\IEEEoverridecommandlockouts

\def\R{\mathbb{R}}

\newcommand{\B}[1]{\boldsymbol{#1}}

\newcommand{\dq}{\mathrm{dq}}

\usepackage{siunitx}
\AtBeginDocument{\DeclareSIUnit{\MWh}{MWh}}
\AtBeginDocument{\DeclareSIUnit{\kWh}{kWh}}
\DeclareSIUnit \voltampere { VA } %apparent power 
\DeclareSIUnit \var { var } %volt-ampere reactive - idle power 
\usepackage{amsmath}
\usepackage{mathtools}
\usepackage{isomath}

\DeclareMathOperator*{\argmax}{arg\,max}
\usepackage{fancyhdr}
\def\BibTeX{{\rm B\kern-.05em{\sc i\kern-.025em b}\kern-.08em
    T\kern-.1667em\lower.7ex\hbox{E}\kern-.125emX}}

\usepackage{titlesec}
\begin{document}

\title{Grid-Forming Vector Current Control FRT Modes Under Symmetrical and Asymmetrical Faults}
\renewcommand{\theenumi}{\alph{enumi}}

\author{
\IEEEauthorblockN{Ognjen Stanojev, Orcun Karaca, Mario Schweizer}%
\IEEEauthorblockA{ABB Corporate Research Center, Switzerland} %
Emails: \{ognjen.stanojev, orcun.karaca, mario.schweizer\}@ch.abb.com
}

\maketitle
\thispagestyle{fancy}
\IEEEpeerreviewmaketitle

%ABSTRACT
\begin{abstract}
Recent research has shown that operating grid-connected converters using the grid-forming vector current control (GFVCC) scheme offers significant benefits, including the simplicity and modularity of the control architecture, as well as enabling a seamless transition from PLL-based grid-following control to grid-forming. An important aspect of any grid-connected converter control strategy is the handling of grid-fault scenarios such as symmetrical and asymmetrical short-circuit faults. This paper presents several fault ride-through (FRT) strategies for GFVCC that enable the converter to provide fault current and stay synchronized to the grid while respecting the converter hardware limitations and retaining grid-forming behavior. The converter control scheme is extended in a modular manner to include negative-sequence loops, and the proposed FRT strategies address both symmetrical and asymmetrical faults. The proposed FRT strategies are analyzed through case studies, including infinite-bus setups and multi-unit grids.
\end{abstract}

%INDEX TERMS

\section{Introduction} \label{sec:intro}

With the increasing integration of power electronic interfaces in power systems, the industry is experiencing a growing demand for grid-forming control strategies. Key characteristics of these control schemes include inertia emulation, power sharing, and voltage magnitude forming. The most extensively analyzed control methods in the literature and industry that can deliver these grid services are droop control~\cite{Yao2017}, power synchronization control~\cite{PSC2010}, virtual synchronous machine control~\cite{VSM2015}, and virtual oscillator control~\cite{VOC2014}. Although numerous studies have demonstrated the effectiveness of these methods, they present several limitations. Key challenges include high implementation complexity and difficulties in parameter tuning stemming from the large number of control parameters, which in turn demand considerable commissioning efforts.

A recently proposed grid-forming strategy, known as grid-forming vector current control (GFVCC), aims to address the limitations of existing methods \cite{Schweizer2022}. This approach combines the functionalities of a virtual synchronous condenser and a virtual current source. The virtual current source employs the standard vector current control \cite{VCC1998} with a phase-locked loop (PLL), while the virtual synchronous condenser provides virtual inertia and enables voltage forming. GFVCC offers several key advantages over traditional grid-forming schemes, including a simplified tuning process due to fewer control parameters, inherent overcurrent handling capabilities, and a modular design. Additionally, it allows for a seamless transition from PLL-based grid-following to grid-forming operation. However, given the recent introduction of the GFVCC concept, fault ride-through (FRT) strategies for symmetrical and asymmetrical faults are yet to be developed for this approach.

The origin of all challenges related to the behavior of grid-forming converters under short-circuit faults lies in the limited overcurrent capabilities of converter semiconductor devices. The tight current limits of converters restrict the grid-forming control capabilities during faults and require altering the control architecture in order to protect the device against thermal hardware damage.
The literature presents a variety of strategies for current limiting in grid-forming converters. These include circular \cite{Sadeghkhani2017} and elliptical \cite{Moawwad2014} current-reference saturation, switch-level current limiting \cite{Du2023}, threshold-based virtual impedance \cite{Paquette2015}, power setpoint modulation \cite{Liu2022}, etc. For a detailed review and comparison of these and other current limiting techniques, the reader is referred to \cite{Baeckeland2024}.

At the same time, a conflicting requirement is imposed by the power system protection, i.e., to provide the maximum possible fault current in order to ensure reliable and timely fault clearing. Moreover, the activation of current limiting mechanisms alters the output behavior of the converter, which can have adverse effects on the rest of the power system. These effects may include increased harmonic distortion in voltages and currents \cite{Du2023}, transient stability (synchronization) issues \cite{Huang2019}, and voltage oscillations \cite{Lin2021}. Therefore, further adjustments to the grid-forming control schemes may be necessary to reconcile the conflicting demands of current limiting and fault current provision while also maintaining transient stability and avoiding other potential adverse effects.

Asymmetrical faults, such as single-phase-to-ground or line-to-line(-to-ground) faults, impose additional challenges that must be addressed in FRT strategies \cite{Rosso2021}. During unbalanced faults, positive-sequence and negative-sequence currents flow through the network, leading to operational challenges such as unbalanced current and voltage waveforms, and oscillations in active and reactive power. To effectively manage these conditions, it is essential to include negative-sequence control loops in the grid-forming design. Notably, grid code and standard requirements \cite{ieee2800} for converter behavior during asymmetrical faults are often defined in terms of negative-sequence voltage and current components. Furthermore, by actively controlling both positive- and negative-sequence components, converters achieve higher controllability over current and voltage waveforms, enabling them to maintain system stability and reduce power quality issues during fault conditions.

Most grid-forming control implementations are not inherently suited for managing fault scenarios, as they fail to meet the requirements outlined above. Consequently, further refinements and enhancements to these control schemes are needed. In this context, the virtual synchronous machine control \cite{VSM2015} has been augmented in \cite{NS-VSM2022} to incorporate negative-sequence control loops. In \cite{DSVOC2023}, a double synchronous unified virtual oscillator controller is proposed to extend the basic virtual oscillator control \cite{VOC2014} to retain synchronization regardless of the nature of grid faults. In \cite{Bhagwat2023}, the droop control method \cite{Yao2017} was enhanced by incorporating individual droop controls for each phase, facilitating phase-specific current limiting under unbalanced fault conditions. However, comparable enhancements have not been explored for GFVCC thus far.

In this paper, we extend the original GFVCC concept by separating the control scheme into positive- and negative-sequence functionalities. The positive-sequence control loops closely follow the original GFVCC design, while various options are explored for the negative-sequence loops, including converter current balancing, power oscillation suppression, and voltage balancing. Additionally, two current limiter designs are proposed to address overcurrent conditions in both balanced and unbalanced fault scenarios. The first approach is based on equal downscaling of the positive- and negative-sequence components, while the second prioritizes the negative-sequence component to ensure that the associated control objectives are met. Building on this foundation, three FRT strategies for GFVCC are proposed that can achieve the given FRT objectives. The properties and implementation requirements of these strategies are analyzed, demonstrating how GFVCC’s modularity allows for structural adaptations under fault scenarios. We showcase the effectiveness of the proposed control scheme by performing numerical simulations on an infinite-bus test system as well as on a larger IEEE network.

The rest of the paper is organized as follows. Firstly, in Sec.~\ref{sec:GFVCC_overview}, we extend the original GFVCC design into a dual-sequence structure. Subsequently, in Sec.~\ref{sec:sym_faults}, we propose three FRT strategies to handle potential issues that can arise under both symmetrical and asymmetrical faults. Specific asymmetrical fault requirements are analyzed separately in Sec.~\ref{sec:asym_faults}, and negative-sequence control loops are proposed. Finally, numerical simulation results are presented in Sec.~\ref{sec:res}.

\textit{Notation.} Parameters are represented by uppercase letters, such as $K$. Bold uppercase letters, like $\boldsymbol{K}$, indicate parameter matrices.
Functions in uppercase letters denote transfer functions, for example $G(s)$.
%, while transfer function matrices are shown in bold, such as $\boldsymbol{G}(s)$. 
The complex frequency variable $s$ can also be interpreted as the derivative operator when appropriate.
Scalar variables are indicated by lowercase letters, e.g., $v$, and their vector forms are represented in bold, such as $\boldsymbol{v}$. We use $\B{v}_\mathrm{abc} \coloneqq [v_\mathrm{a}~v_\mathrm{b}~v_\mathrm{c}]^\top$ and $\B{i}_\mathrm{abc} \coloneqq [i_\mathrm{a}~i_\mathrm{b}~i_\mathrm{c}]^\top$ to denote three-phase voltage and current quantities in their natural $\mathrm{abc}$ form; $\B{v}\coloneqq v_\alpha + jv_\beta$ and $\B{i}\coloneqq i_\alpha + ji_\beta$ to denote voltage and current vectors in the stationary reference frame; $\B{v}_\dq\coloneqq v_\mathrm{d} + jv_\mathrm{q}$ and $\B{i}_\dq \coloneqq i_\mathrm{d} + ji_\mathrm{q}$ to denote voltage and current vectors in the rotational reference frame. Depending on the context, these variables can also be interpreted in vector form, e.g., $\B{v}\coloneqq[v_\alpha~v_\beta]^\top$ or  $\B{v}_\dq\coloneqq[v_\mathrm{d}~v_\mathrm{q}]^\top$. Subscripts ${}_+$ and ${}_-$, indicate positive- and negative-sequence components. Transformation of three-phase to $\dq$ quantities is achieved via
\begin{equation*}
	\B{P}(\theta) = \frac{2}{3}\begin{bmatrix}
		\cos(\theta) & \cos(\theta - \frac{2\pi}{3}) & \cos(\theta + \frac{2\pi}{3})  \\
		-\sin(\theta) & -\sin(\theta - \frac{2\pi}{3}) & -\sin(\theta + \frac{2\pi}{3})
	\end{bmatrix},
\end{equation*}
with $\theta$ denoting the angle of rotation of the reference frame. Given an angle $\theta$, the 2-D rotation matrix is given by
\begin{equation*}
    \B{R}(\theta) \coloneqq \begin{bmatrix}\cos{\theta} & -\sin{\theta}\\  \sin{\theta} & \cos{\theta} \end{bmatrix} \in \R^{2 \times 2}.
\end{equation*}

\section{Grid-Forming Vector Current Control Including Negative-Sequence Control Loops} \label{sec:GFVCC_overview}
In this section, we extend the GFVCC concept into a dual-sequence structure, in which the positive-sequence control loops closely follow the original GFVCC design and the negative-sequence loops are introduced to govern the converter response under unbalanced conditions. 
For more information on the basic GFVCC concept derivation, tuning guidelines, and experimental verification, we refer the reader to the original paper~\cite{Schweizer2022}. 

\subsection{Converter System}
\begin{figure}[!b]
	\centering
	\includegraphics[scale=0.75]{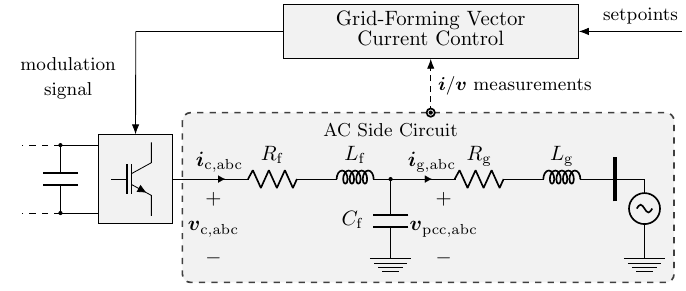}
	\caption{Single-line representation of the considered converter system in the natural $\mathrm{abc}$ reference frame.}
	\label{fig:system-model}
\end{figure}

An overview of the considered system is depicted in Fig.~\ref{fig:system-model}, where a single-line diagram of an inverter connected to a grid equivalent through an RLC filter is shown. 
The power electronic system considered in this paper consists of an ideal DC voltage source, a lossless DC/AC converter, and an RLC filter $(R_\mathrm{f},L_\mathrm{f},C_\mathrm{f})$ on the converter AC side. The grid equivalent impedance consists of an (typically unknown) inductor $L_\mathrm{g}$ and a resistor $R_\mathrm{g}$. The DC side dynamics are not considered as the focus is on the system-level converter controls, and a separate control loop is assumed to keep the DC-link voltage stable. Voltage and current at the converter terminal (output) are denoted with subscript $\mathrm{c}$, i.e., respectively by $\boldsymbol{v}_\mathrm{c,abc}$ and $\boldsymbol{i}_\mathrm{c,abc}$. The point of common coupling (PCC) voltage is denoted by $\B{v}_\mathrm{pcc,abc}$, whereas the current injected into the grid equivalent is designated by $\B{i}_\mathrm{g,abc}$. The converter is controlled using the extended GFVCC scheme, which from a high-level perspective takes voltage and power setpoints and voltage and current measurements as inputs and generates a modulation signal for the switching stage as the output.

\subsection{Sequence Extraction Process}
Asymmetrical faults cause the three-phase voltages and currents $\boldsymbol{\xi}_\mathrm{abc}\in\{\boldsymbol{v}_\mathrm{abc},\boldsymbol{i}_\mathrm{abc}\}$ to become unbalanced. Under these circumstances, access to the positive- and negative-sequence components of the voltages and currents $\boldsymbol{\xi}_\mathrm{abc} = \boldsymbol{\xi}_\mathrm{abc+}+\boldsymbol{\xi}_\mathrm{abc-}$ is needed to enable shaping of the three-phase signals at the converter output. Simply applying the Park transform to $\boldsymbol{\xi}_\mathrm{abc}$ at the fundamental frequency in positive $\B{P}(+\omega t)$ and negative $\B{P}(-\omega t)$ directions does not provide the desired result due to the appearance of oscillatory cross-coupling terms:
 \begin{align*}
     \B{P}(+\omega t)\boldsymbol{\xi}_\mathrm{abc} &= \boldsymbol{\xi}_{\dq+} + \B{R}(2\omega t)\boldsymbol{\xi}_\mathrm{dq-}, \\
     \B{P}(-\omega t)\boldsymbol{\xi}_\mathrm{abc} &= \boldsymbol{\xi}_{\dq-} + \B{R}(-2\omega t)\boldsymbol{\xi}_\mathrm{dq+}.
 \end{align*}
 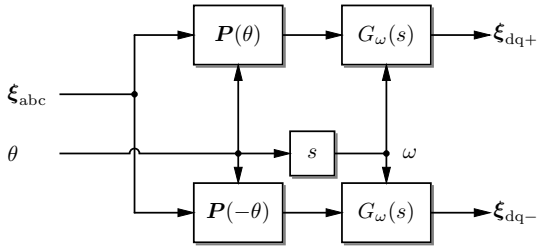
\begin{figure}[!b]
	\centering
	\resizebox{0.4\textwidth}{!}{\begin{tikzpicture}[scale=2.54]%
% dpic version 2024.01.01 option -g for TikZ and PGF 1.01
\ifx\dpiclw\undefined\newdimen\dpiclw\fi
\global\def\dpicdraw{\draw[line width=\dpiclw]}
\global\def\dpicstop{;}
\dpiclw=0.8bp
\dpiclw=0.8bp
\draw (0,0) node(I1)[right=-2bp]{$\boldsymbol{\xi}_\mathrm{abc}$};
\dpicdraw (0.35,0)
 --(0.825,0)\dpicstop
\dpicdraw[fill=black](0.825,0) circle (0.005906in)\dpicstop
\dpicdraw (0.825,0)
 --(0.825,0.375)\dpicstop
\dpicdraw[line width=0.4bp](0.825,0.375) circle (0.00109in)\dpicstop
\filldraw[line width=0bp](1.1,0.35)
 --(1.2,0.375)
 --(1.1,0.4) --cycle\dpicstop
\dpicdraw (0.825,0.375)
 --(1.177094,0.375)\dpicstop
\dpicdraw (1.2,0.1875) rectangle (1.7625,0.5625)\dpicstop
\draw (1.48125,0.375) node{$\B{P}(\theta)$};
\dpicdraw[line width=1bp,draw=gray](1.2125,0.175)
 --(1.775,0.175)
 --(1.775,0.55)\dpicstop
\filldraw[line width=0bp](2.0375,0.35)
 --(2.1375,0.375)
 --(2.0375,0.4) --cycle\dpicstop
\dpicdraw (1.7625,0.375)
 --(2.114594,0.375)\dpicstop
\dpicdraw (2.1375,0.1875) rectangle (2.7,0.5625)\dpicstop
\draw (2.41875,0.375) node{$G_\omega(s)$};
\dpicdraw[line width=1bp,draw=gray](2.15,0.175)
 --(2.7125,0.175)
 --(2.7125,0.55)\dpicstop
\filldraw[line width=0bp](2.975,0.35)
 --(3.075,0.375)
 --(2.975,0.4) --cycle\dpicstop
\dpicdraw (2.7,0.375)
 --(3.052094,0.375)\dpicstop
\draw (3.075,0.375) node[right=-2bp]{$\boldsymbol{\xi}_\mathrm{dq+}$};
\dpicdraw (0.825,0)
 --(0.825,-0.75)\dpicstop
\dpicdraw[line width=0.4bp](0.825,-0.75) circle (0.00109in)\dpicstop
\filldraw[line width=0bp](1.1,-0.775)
 --(1.2,-0.75)
 --(1.1,-0.725) --cycle\dpicstop
\dpicdraw (0.825,-0.75)
 --(1.177094,-0.75)\dpicstop
\dpicdraw (1.2,-0.9375) rectangle (1.7625,-0.5625)\dpicstop
\draw (1.48125,-0.75) node{$\B{P}(-\theta)$};
\dpicdraw[line width=1bp,draw=gray](1.2125,-0.95)
 --(1.775,-0.95)
 --(1.775,-0.575)\dpicstop
\filldraw[line width=0bp](2.0375,-0.775)
 --(2.1375,-0.75)
 --(2.0375,-0.725) --cycle\dpicstop
\dpicdraw (1.7625,-0.75)
 --(2.114594,-0.75)\dpicstop
\dpicdraw (2.1375,-0.9375) rectangle (2.7,-0.5625)\dpicstop
\draw (2.41875,-0.75) node{$G_\omega(s)$};
\dpicdraw[line width=1bp,draw=gray](2.15,-0.95)
 --(2.7125,-0.95)
 --(2.7125,-0.575)\dpicstop
\filldraw[line width=0bp](2.975,-0.775)
 --(3.075,-0.75)
 --(2.975,-0.725) --cycle\dpicstop
\dpicdraw (2.7,-0.75)
 --(3.052094,-0.75)\dpicstop
\draw (3.075,-0.75) node[right=-2bp]{$\boldsymbol{\xi}_\mathrm{dq-}$};
\draw (0,-0.375) node[right=-2bp]{$\theta$};
\dpicdraw (0.35,-0.375)
 --(0.799306,-0.375)\dpicstop
\dpicdraw (0.79375,-0.375)
 ..controls (0.79375,-0.363835) and (0.799706,-0.353519)
 ..(0.809375,-0.347937)
 ..controls (0.819044,-0.342354) and (0.830956,-0.342354)
 ..(0.840625,-0.347937)
 ..controls (0.850294,-0.353519) and (0.85625,-0.363835)
 ..(0.85625,-0.375)\dpicstop
\dpicdraw (0.850694,-0.375)
 --(1.48125,-0.375)\dpicstop
\dpicdraw[fill=black](1.48125,-0.375) circle (0.005906in)\dpicstop
\filldraw[line width=0bp](1.50625,0.0875)
 --(1.48125,0.1875)
 --(1.45625,0.0875) --cycle\dpicstop
\dpicdraw (1.48125,-0.375)
 --(1.48125,0.164594)\dpicstop
\filldraw[line width=0bp](1.45625,-0.4625)
 --(1.48125,-0.5625)
 --(1.50625,-0.4625) --cycle\dpicstop
\dpicdraw (1.48125,-0.375)
 --(1.48125,-0.539594)\dpicstop
\filldraw[line width=0bp](1.709375,-0.4)
 --(1.809375,-0.375)
 --(1.709375,-0.35) --cycle\dpicstop
\dpicdraw (1.48125,-0.375)
 --(1.786469,-0.375)\dpicstop
\dpicdraw (1.809375,-0.515625) rectangle (2.090625,-0.234375)\dpicstop
\draw (1.95,-0.375) node{$s$};
\dpicdraw[line width=1bp,draw=gray](1.821875,-0.528125)
 --(2.103125,-0.528125)
 --(2.103125,-0.246875)\dpicstop
\dpicdraw (2.090625,-0.375)
 --(2.41875,-0.375)\dpicstop
\dpicdraw[fill=black](2.41875,-0.375) circle (0.005906in)\dpicstop
\filldraw[line width=0bp](2.44375,0.0875)
 --(2.41875,0.1875)
 --(2.39375,0.0875) --cycle\dpicstop
\dpicdraw (2.41875,-0.375)
 --(2.41875,0.164594)\dpicstop
\filldraw[line width=0bp](2.39375,-0.4625)
 --(2.41875,-0.5625)
 --(2.44375,-0.4625) --cycle\dpicstop
\dpicdraw (2.41875,-0.375)
 --(2.41875,-0.539594)\dpicstop
\draw (2.56875,-0.375) node{$\omega$};
\end{tikzpicture}%
}
	\caption{The process of positive $\B{\xi}_{\dq+}$ and negative $\B{\xi}_\mathrm{dq-}$ sequence extraction from a three-phase quantity $\B{\xi}_\mathrm{abc}\in\{v_\mathrm{abc},i_\mathrm{abc}\}$ in the natural reference frame and a given reference angle $\theta$.}
	\label{fig:sequence_extraction}
\end{figure}

These oscillations cause tracking difficulties for the inner control loops of converters, especially in the case of medium voltage high power converters operating at relatively low switching frequencies. 
Therefore, it is desirable to filter out these oscillations to facilitate control of the injected currents under unbalanced conditions. One of the most straightforward solutions to attenuate the $2\omega$ oscillations is to use notch filters as depicted in Fig.~\ref{fig:sequence_extraction}. The filter is designed as an adaptive biquad filter, given by the following transfer function:
\begin{equation*}
	G_\omega(s) = \frac{s^2 +4\omega^2}{s^2+4\zeta\omega s+4\omega^2},
\end{equation*}
where $\zeta$ is the filter damping factor, and the central rejected frequency of the filter is adaptively selected in real-time based on the frequency input $\omega$ to ensure proper operation even under large frequency deviations in the grid. A similar performance can be achieved with a decoupling network \cite{teodorescu_grid_2011} or using other sequence extraction methods \cite{AwalSequences2022}.

\subsection{GFVCC Overview}
\begin{figure}[!t]
	\resizebox{0.485\textwidth}{!}{\begin{tikzpicture}[scale=2.54]%
% dpic version 2024.01.01 option -g for TikZ and PGF 1.01
\ifx\dpiclw\undefined\newdimen\dpiclw\fi
\global\def\dpicdraw{\draw[line width=\dpiclw]}
\global\def\dpicstop{;}
\dpiclw=0.8bp
\dpiclw=0.8bp
\dpicdraw (0,-0.28125) rectangle (0.84375,0.28125)\dpicstop
\draw (0.421875,0) node{\shortstack{Virtual\\%
Synchronous\\%
Condenser}};
\dpicdraw[line width=1bp,draw=gray](0.0125,-0.29375)
 --(0.85625,-0.29375)
 --(0.85625,0.26875)\dpicstop
\filldraw[line width=0bp](-0.1,0.115625)
 --(0,0.140625)
 --(-0.1,0.165625) --cycle\dpicstop
\dpicdraw (-0.5625,0.140625)
 --(-0.022906,0.140625)\dpicstop
\draw (-0.292703,0.140625) node[left=-2bp]{$\boldsymbol{v}_{\mathrm{pcc},\mathrm{dq+}}\qquad$};
\filldraw[line width=0bp](-0.1,-0.165625)
 --(0,-0.140625)
 --(-0.1,-0.115625) --cycle\dpicstop
\dpicdraw (-0.5625,-0.140625)
 --(-0.022906,-0.140625)\dpicstop
\draw (-0.292703,-0.140625) node[left=-2bp]{$\boldsymbol{v}_{\mathrm{set},\mathrm{dq+}}\qquad$};
\dpicdraw (0,-1.125) rectangle (0.84375,-0.5625)\dpicstop
\draw (0.421875,-0.84375) node{\shortstack{Virtual\\%
Current\\%
Source}};
\dpicdraw[line width=1bp,draw=gray](0.0125,-1.1375)
 --(0.85625,-1.1375)
 --(0.85625,-0.575)\dpicstop
\dpicdraw[fill=black](-0.375,-0.84375) circle (0.005906in)\dpicstop
\dpicdraw[fill=black](-0.1875,-0.140625) circle (0.005906in)\dpicstop
\filldraw[line width=0bp](-0.1,-0.68125)
 --(0,-0.65625)
 --(-0.1,-0.63125) --cycle\dpicstop
\dpicdraw (-0.1875,-0.140625)
 --(-0.1875,-0.65625)
 --(-0.022906,-0.65625)\dpicstop
\dpicdraw[fill=black](-0.375,0.140625) circle (0.005906in)\dpicstop
\dpicdraw (-0.375,0.140625)
 --(-0.375,-0.114931)\dpicstop
\dpicdraw (-0.375,-0.109375)
 ..controls (-0.357741,-0.109375) and (-0.34375,-0.123366)
 ..(-0.34375,-0.140625)
 ..controls (-0.34375,-0.157884) and (-0.357741,-0.171875)
 ..(-0.375,-0.171875)\dpicstop
\dpicdraw (-0.375,-0.166319)
 --(-0.375,-0.84375)\dpicstop
\filldraw[line width=0bp](-0.1,-0.86875)
 --(0,-0.84375)
 --(-0.1,-0.81875) --cycle\dpicstop
\dpicdraw (-0.375,-0.84375)
 --(-0.022906,-0.84375)\dpicstop
\filldraw[line width=0bp](-0.1,-1.05625)
 --(0,-1.03125)
 --(-0.1,-1.00625) --cycle\dpicstop
\dpicdraw (-0.5625,-1.03125)
 --(-0.022906,-1.03125)\dpicstop
\draw (-0.292703,-1.03125) node[left=-2bp]{$\boldsymbol{s}_{\mathrm{set}}\qquad$  };
\filldraw[line width=0bp](0.396875,-0.4625)
 --(0.421875,-0.5625)
 --(0.446875,-0.4625) --cycle\dpicstop
\dpicdraw (0.421875,-0.28125)
 --(0.421875,-0.539594)\dpicstop
\draw (0.421875,-0.410422) node[left=-2bp]{$\Delta\omega_\mathrm{r}$};
\dpicdraw (0,-1.96875) rectangle (0.84375,-1.40625)\dpicstop
\draw (0.421875,-1.6875) node{\shortstack{Negative\\%
Sequence\\%
Control}};
\dpicdraw[line width=1bp,draw=gray](0.0125,-1.98125)
 --(0.85625,-1.98125)
 --(0.85625,-1.41875)\dpicstop
\dpicdraw (-0.375,-0.84375)
 --(-0.375,-1.005556)\dpicstop
\dpicdraw (-0.375,-1)
 ..controls (-0.357741,-1) and (-0.34375,-1.013991)
 ..(-0.34375,-1.03125)
 ..controls (-0.34375,-1.048509) and (-0.357741,-1.0625)
 ..(-0.375,-1.0625)\dpicstop
\dpicdraw (-0.375,-1.056944)
 --(-0.375,-1.5)\dpicstop
\dpicdraw[line width=0.4bp](-0.375,-1.5) circle (0.00109in)\dpicstop
\filldraw[line width=0bp](-0.1,-1.525)
 --(0,-1.5)
 --(-0.1,-1.475) --cycle\dpicstop
\dpicdraw (-0.375,-1.5)
 --(-0.022906,-1.5)\dpicstop
\filldraw[line width=0bp](-0.1,-1.7125)
 --(0,-1.6875)
 --(-0.1,-1.6625) --cycle\dpicstop
\dpicdraw (-0.5625,-1.6875)
 --(-0.022906,-1.6875)\dpicstop
\draw (-0.292703,-1.6875) node[left=-2bp]{$\boldsymbol{v}_{\mathrm{pcc},\mathrm{dq-}}\qquad$  };
\filldraw[line width=0bp](-0.1,-1.9)
 --(0,-1.875)
 --(-0.1,-1.85) --cycle\dpicstop
\dpicdraw (-0.5625,-1.875)
 --(-0.022906,-1.875)\dpicstop
\draw (-0.292703,-1.875) node[left=-2bp]{$\boldsymbol{v}_{\mathrm{set},\mathrm{dq-}}\qquad$  };
\filldraw[line width=0bp](1.30625,-0.86875)
 --(1.40625,-0.84375)
 --(1.30625,-0.81875) --cycle\dpicstop
\dpicdraw (0.84375,-0.84375)
 --(1.383344,-0.84375)\dpicstop
\draw (1.113547,-0.84375) node[above=-2bp]{$\boldsymbol{i}_{\mathrm{vcs},\mathrm{dq}+}$};
\dpicdraw (1.49375,-0.84375) circle (0.034449in)\dpicstop
\draw (1.49375,-0.84375) node{+};
\filldraw[line width=0bp](1.46875,-0.65625)
 --(1.49375,-0.75625)
 --(1.51875,-0.65625) --cycle\dpicstop
\dpicdraw (0.84375,0)
 --(1.49375,0)
 --(1.49375,-0.733344)\dpicstop
\draw (1.16875,0) node[above=-2bp]{$\boldsymbol{i}_{\mathrm{vsc},\mathrm{dq}+}$};
\dpicdraw (2.0625,-1.828125) rectangle (2.90625,-1.265625)\dpicstop
\draw (2.484375,-1.546875) node{\shortstack{Current\\%
Limiter}};
\dpicdraw[line width=1bp,draw=gray](2.075,-1.840625)
 --(2.91875,-1.840625)
 --(2.91875,-1.278125)\dpicstop
\filldraw[line width=0bp](1.9625,-1.43125)
 --(2.0625,-1.40625)
 --(1.9625,-1.38125) --cycle\dpicstop
\dpicdraw (1.49375,-0.93125)
 --(1.49375,-1.40625)
 --(2.039594,-1.40625)\dpicstop
\draw (1.766672,-1.40625) node[above=-2bp]{$\boldsymbol{i}_{\mathrm{dq}+}^\mathrm{ref}$};
\dpicdraw[fill=black](1.49375,-1.20625) circle (0.005906in)\dpicstop
\filldraw[line width=0bp](0.396875,-1.30625)
 --(0.421875,-1.40625)
 --(0.446875,-1.30625) --cycle\dpicstop
\dpicdraw (1.49375,-1.20625)
 --(0.421875,-1.20625)
 --(0.421875,-1.383344)\dpicstop
\filldraw[line width=0bp](1.9625,-1.7125)
 --(2.0625,-1.6875)
 --(1.9625,-1.6625) --cycle\dpicstop
\dpicdraw (0.84375,-1.6875)
 --(1.49375,-1.6875)
 --(1.49375,-1.6875)
 --(2.039594,-1.6875)\dpicstop
\draw (1.766672,-1.6875) node[below=-2bp]{$\boldsymbol{i}_{\mathrm{dq}-}^\mathrm{ref}$};
\dpicdraw (3.421875,-1.828125) rectangle (4.265625,-1.265625)\dpicstop
\draw (3.84375,-1.546875) node{\shortstack{Current\\%
Control}};
\dpicdraw[line width=1bp,draw=gray](3.434375,-1.840625)
 --(4.278125,-1.840625)
 --(4.278125,-1.278125)\dpicstop
\filldraw[line width=0bp](3.321875,-1.43125)
 --(3.421875,-1.40625)
 --(3.321875,-1.38125) --cycle\dpicstop
\dpicdraw (2.90625,-1.40625)
 --(3.398969,-1.40625)\dpicstop
\draw (3.152609,-1.40625) node[above=-2bp]{$\bar{\boldsymbol{i}}_{\mathrm{dq}+}^\mathrm{ref}$};
\filldraw[line width=0bp](3.321875,-1.7125)
 --(3.421875,-1.6875)
 --(3.321875,-1.6625) --cycle\dpicstop
\dpicdraw (2.90625,-1.6875)
 --(3.398969,-1.6875)\dpicstop
\draw (3.152609,-1.6875) node[below=-2bp]{$\bar{\boldsymbol{i}}_{\mathrm{dq}-}^\mathrm{ref}$};
\filldraw[line width=0bp](4.540625,-1.571875)
 --(4.640625,-1.546875)
 --(4.540625,-1.521875) --cycle\dpicstop
\dpicdraw (4.265625,-1.546875)
 --(4.617719,-1.546875)\dpicstop
\draw (4.441672,-1.546875) node[above=-2bp]{$\boldsymbol{v}_\mathrm{ref}$};
\filldraw[line width=0bp](4.029688,-1.165625)
 --(4.054688,-1.265625)
 --(4.079687,-1.165625) --cycle\dpicstop
\dpicdraw (4.054688,-1.242719)
 --(4.054688,-0.890625)\dpicstop
\draw (4.054688,-0.790625) node{$\B{v}_\mathrm{pcc}$};
\filldraw[line width=0bp](3.607812,-1.165625)
 --(3.632813,-1.265625)
 --(3.657813,-1.165625) --cycle\dpicstop
\dpicdraw (3.632813,-1.242719)
 --(3.632813,-0.890625)\dpicstop
\draw (3.632813,-0.790625) node{$\B{i}_\mathrm{c}$};
\end{tikzpicture}%
}
	\caption{Extended GFVCC structure including both positive- and negative-sequence controls. The overall controller takes voltage and power setpoints as control inputs and outputs voltage reference $\boldsymbol{v}_\mathrm{ref}$ for the switching stage.}
	\label{fig:gfvcc_overview}
\end{figure}
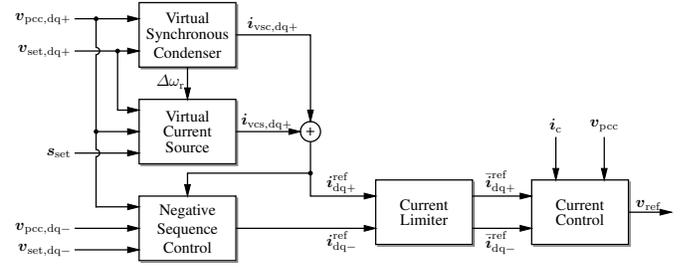

A control block diagram giving a general overview of the extended GFVCC structure is depicted in Fig.~\ref{fig:gfvcc_overview}. The diagram includes both the original GFVCC loops, cast here in positive-sequence, as well as the negative-sequence loops introduced in this paper. Note that the input measurement signals indicated in the positive- or negative-sequence $\dq$ domain are obtained by processing the corresponding $\mathrm{abc}$ quantities according to the sequence extraction process given in Fig.~\ref{fig:sequence_extraction}.

The positive-sequence controls closely follow the control design proposed in \cite{Schweizer2022}, where the grid-forming functionality is separated into two parts: a virtual synchronous condenser and a virtual current source. The behavior of the virtual synchronous condenser is emulated with PLL and a virtual admittance block based on PCC positive-sequence voltage measurement $\B{v}_{\mathrm{pcc},\dq+}$ and voltage setpoint $\B{v}_{\mathrm{set},\dq+}$. 
The virtual current source is responsible for power setpoint $\B{s}_\mathrm{set}\coloneqq[P_\mathrm{set}~Q_\mathrm{set}]^\top$ tracking and acts simultaneously as a governor based on the frequency deviation input $\Delta\omega_\mathrm{r}$. 
The current references from the two functional parts, $\B{i}_{\mathrm{vsc},\dq+}$ and $\B{i}_{\mathrm{vcs},\dq+}$, are generated separately and then superimposed to construct a total positive-sequence reference $\B{i}_{\dq+}^\mathrm{ref}$ sent to the current limiter block. 

The considered negative-sequence loops define the converter response to unbalanced voltages measured at its terminals and will be thoroughly discussed in Sec.~\ref{sec:asym_faults} together with different control goals and controller implementations. Both positive- and negative-sequence current references are sent to the current limiter, where they are potentially adjusted to keep the current flow through the semiconductors within the allowable range. The current limiter implementation is discussed in detail in Sec.~\ref{sec:current_limiter}. Finally, the current control loop enforces the current reference by generating a voltage reference based on a stationary frame proportional controller with the filtered PCC voltage feedforward, as discussed in Sec.~\ref{sec:current_control}.

\subsubsection{Virtual Synchronous Condenser}
The virtual synchronous condenser part of the control scheme serves two primary functions: (i) facilitating synchronization and providing virtual inertia and damper winding emulation to the system and (ii) forming voltage. These objectives are achieved through the combined action of the PLL and virtual admittance. The typical type-2 PLL implementation is employed, which estimates the phase angle $\theta_\mathrm{r}$ and the grid frequency $\omega_\mathrm{r}$ using a PI controller ($K_\mathrm{pll,p},K_\mathrm{pll,i}$) that diminishes the q-component of the positive-sequence PCC voltage $\B{v}_{\mathrm{pcc},\mathrm{dq}+}$, as follows:
\begin{align}
    \omega_\mathrm{r} &= \omega_\mathrm{n} + \left( K_\mathrm{pll,p} + \frac{K_\mathrm{pll,i}}{s} \right) v_{\mathrm{pcc},\mathrm{q}+},\label{eq:PLL} \\
    \theta_\mathrm{r} &= \frac{1}{s}\omega_\mathrm{r},
\end{align}
where $\omega_\mathrm{n}$ is the nominal angular frequency of the system. A virtual admittance defined by virtual resistance $R_\mathrm{v}$ and virtual inductance $L_\mathrm{v}$ is implemented as 
\begin{equation}\label{eq:virtual_admittance}
    \B{i}_{\mathrm{vsc},\dq+}=\B{Y}_\mathrm{v}(s)(\B{v}_{\mathrm{set},\dq+}-\B{v}_{\mathrm{pcc},\mathrm{dq}+}),
\end{equation} 
where $\B{Y}_\mathrm{v}(s)=(R_\mathrm{v}+(s+j\omega_\mathrm{n})L_\mathrm{v})^{-1}$. The setpoint voltage vector is defined as $\B{v}_{\mathrm{set},\dq+} = V_\mathrm{v}\angle{\theta_\mathrm{r}}$, thereby providing voltage forming due to constant magnitude $V_\mathrm{v}$.

Selecting the PLL gains in the following fashion:
\begin{equation} \label{eq:PLL_gains}
	K_\mathrm{pll,p} = \frac{K_\mathrm{f}}{M \norm{\B{v}_{\mathrm{pcc},\mathrm{dq}+}}},\,\, K_\mathrm{pll,i} = \frac{V_\mathrm{v}}{M X_\mathrm{v}},\,\, X_\mathrm{v} = \omega_\mathrm{n}L_\mathrm{v}, 
\end{equation}
leads to the emulation of inertia and damping with respective constants $M$ and $K_\mathrm{f}$, equivalently to virtual synchronous machine grid-forming controls \cite{VSM2007}. Instead of using $\norm{\B{v}_{\mathrm{pcc},\mathrm{dq}+}}$ in the denominator, we use a constant that corresponds to the nominal voltage for simplicity. The equivalence can be demonstrated by assuming $R_\mathrm{v}\approx0$, substituting \eqref{eq:PLL_gains} in \eqref{eq:PLL}, and taking a derivative of thus obtained equation, which results in
\begin{equation}\label{eq:vsm_equiv}
    sM\omega_r = -\frac{V_\mathrm{v}\norm{\B{v}_{\mathrm{pcc},\mathrm{dq}+}}}{X_\mathrm{v}}\sin(\theta_\mathrm{r}-\theta_\mathrm{pcc}) - K_\mathrm{f}(\omega_\mathrm{r}-\omega_\mathrm{pcc}),
\end{equation}
where the first term reflects the power flowing through the virtual admittance and $\omega_\mathrm{pcc}=s\theta_\mathrm{pcc}$ is the angular frequency of PCC voltage. For a more detailed derivation, refer to \cite{Schweizer2022}.

\subsubsection{Virtual Current Source}
The virtual current source consists of three parts: (i) a governor, (ii) an automatic voltage regulator (AVR), and, if necessary, (iii) an active damping resistor. 
The governor's function of maintaining power balance in steady-state is achieved through a droop control loop, which determines the d-axis component of the virtual current source:
\begin{equation}\label{eq:governor}
    i_\mathrm{v,d+} = \frac{P_\mathrm{set}+K_\mathrm{g}G_\mathrm{lpf}^\mathrm{f}(s)(\omega_r-\omega_\mathrm{n})}{G_\mathrm{lpf}^\mathrm{v}(s)v_\mathrm{pcc,d+}},
\end{equation}
where $K_\mathrm{g}$ is the droop gain, and $G_\mathrm{lpf}^\mathrm{f}(s)$ and $G_\mathrm{lpf}^\mathrm{v}(s)$ are low-pass filters applied to frequency and voltage measurements.

An AVR is implemented to impose voltage magnitude regulation by impressing a q-component current reference:
\begin{equation}\label{eq:AVR}
    i_\mathrm{v,q+} = K_\mathrm{V}\frac{1}{s}(G_\mathrm{lpf}^\mathrm{v}(s)\norm{\B{v}_{\mathrm{pcc},\mathrm{dq}+}} - V_\mathrm{v}),
\end{equation}
where $K_\mathrm{V}$ is the integral gain. If desired, reactive power setpoint $Q_\mathrm{set}$ tracking can be included as part of the regulator.  

Moreover, the inclusion of a virtual termination resistor within the configuration can enhance stabilization through the implementation of an additional current reference:
\begin{equation}\label{eq:active_daming_pos}
	\boldsymbol{i}_{\mathrm{ad},\dq+} = \frac{1}{R_\mathrm{ad}} {G}_\mathrm{hp}(s)\B{v}_{\mathrm{pcc},\dq+},
\end{equation}
where a high-pass filter ${G}_\mathrm{hp}(s)$ is applied to the PCC voltage measurement, and through a virtual damping resistor $R_\mathrm{ad}$, it generates the reference signal.

Finally, the total current reference of the virtual current source is obtained by adding the individual components as $\B{i}_{\mathrm{vcs},\dq+} = \boldsymbol{i}_{\mathrm{ad},\dq+} + {i}_{\mathrm{v,d+}} + j{i}_{\mathrm{v,q+}}$. Furthermore, individual references from the virtual current source and the virtual synchronous condenser are added to generate the positive-sequence current reference $\boldsymbol{i}_\mathrm{dq+}^\mathrm{ref}=\B{i}_{\mathrm{vcs},\dq+} + \B{i}_{\mathrm{vsc},\dq+}$. As a result, the complete power flowing into the grid in steady state is injected by the current source, while the virtual synchronous condenser injects power only during transients.  

\subsection{Current Limiter}\label{sec:current_limiter}
The current limiter is an integral part of inner control loops, as depicted in Fig.~\ref{fig:gfvcc_overview}. Its main purpose is to ensure that the joint current reference generated by the virtual synchronous condenser, the virtual current source, and the negative-sequence control loops respect the current limits of the converter.\footnote{Apart from ensuring that the current limits are respected, additional requirements can be included such as the reverse power limit.} It is typically implemented as a projection operator, transforming the original current reference into a saturated reference respecting converter limits. The projection operation can be implemented in several ways \cite{BoFan2022} and has a higher complexity when the converter is operating under unbalanced conditions. In the rest of the subsection, we first present the current limiter for balanced conditions and subsequently discuss how it needs to be augmented under unbalanced conditions.

\subsubsection{Balanced Conditions}
Let $\B{i}_{\dq}^{\mathrm{ref}}$ denote the current reference that is passed on to the current limiter. Considering that balanced grid conditions are studied, the current reference under study corresponds to the positive-sequence reference $\B{i}_{\dq}^{\mathrm{ref}}=\B{i}_{\dq+}^{\mathrm{ref}}$. In case the current reference magnitude exceeds the allowable converter current $I_\mathrm{lim}$, i.e.,  $||\B{i}_{\dq}^{\mathrm{ref}}||\geq I_\mathrm{lim}$, the reference needs to be adjusted in order to prevent damaging of the semiconductor devices. Depending on how the saturated current reference is computed, several current limiters have been proposed for grid-forming converters \cite{BoFan2022}, including the instantaneous limiter, the magnitude limiter, and the priority-based limiter. In this paper, we employ the magnitude-based limiter, which only adjusts the reference magnitude while preserving the original angle, more precisely,
\begin{equation}\label{eq:balanced_limiter}
	\B{\bar{i}}_{\dq}^{\mathrm{ref}} = \begin{cases}
		\B{i}_{\dq}^{\mathrm{ref}}, & ||\B{i}_{\dq}^{\mathrm{ref}}||\leq I_\mathrm{lim},\\
		I_\mathrm{lim}\frac{\B{i}_{\dq}^{\mathrm{ref}}}{||\B{i}_{\dq}^{\mathrm{ref}}||}, & ||\B{i}_{\dq}^{\mathrm{ref}}|| > I_\mathrm{lim},
	\end{cases}
\end{equation}
where $\B{\bar{i}}_{\dq}^{\mathrm{ref}}$ denotes the saturated current reference.

\subsubsection{Unbalanced Conditions}
Let $\B{i}_{\dq+}^{\mathrm{ref}}$ denote the positive-sequence current reference and $\B{i}_{\dq-}^{\mathrm{ref}}$ the negative-sequence current reference generated by the outer control loops. Considering that in unbalanced conditions three-phase currents have different magnitudes, the objective of the current limiter is to consider phase current magnitudes of all phases and keep them below the allowable limit $I_\mathrm{lim}$. Magnitudes of the phase currents can be determined based on the $\dq$ references as \cite{DSVOC2023}:
\begin{equation*}
    I_{\mathrm{max},x} = \sqrt{\norm{\B{i}_{\dq+}^{\mathrm{ref}}}^2+\norm{\B{i}_{\dq-}^{\mathrm{ref}}}^2+2\mathrm{Re}\{\B{i}_{\dq+}^{\mathrm{ref}}\B{i}_{\dq-}^{\mathrm{ref}}e^{j2\lambda_x}\}},
\end{equation*}
with $x\in\{\mathrm{a},\mathrm{b},\mathrm{c}\}$ and $\lambda_x\in\{0,-2\pi/3,2\pi/3\}$.
Two methods to achieve the goal are considered in this paper: (i) equal downscaling of positive- and negative-sequence references and (ii) negative-sequence priority downscaling. In the first method, current limiting is achieved by reducing both sequence references with the same factor, as $\B{\bar{i}}_{\dq+}^{\mathrm{ref}} = \gamma \B{{i}}_{\dq+}^{\mathrm{ref}}, \B{\bar{i}}_{\dq-}^{\mathrm{ref}} = \gamma\B{{i}}_{\dq-}^{\mathrm{ref}}$, with the scaling factor defined by
\begin{equation*}
	\gamma = \frac{I_\mathrm{lim}}{\max\{I_{\mathrm{max},\mathrm{a}},I_{\mathrm{max},\mathrm{b}},I_{\mathrm{max},\mathrm{c}}\}},
\end{equation*} 
if $\max\{I_{\mathrm{max},\mathrm{a}},I_{\mathrm{max},\mathrm{b}},I_{\mathrm{max},\mathrm{c}}\}\geq I_\mathrm{lim}$, otherwise, $\gamma=1$.
This approach has the advantage of giving equal priority to both references and, therefore, equal priority to the control goals posed for each sequence controller. 

However, as will be demonstrated in the results section, the positive-sequence reference is naturally higher in magnitude as it is associated with voltage forming and synchronization.
This might lead to only marginal participation of the negative-sequence reference in the impressed current vector and failure to achieve certain negative-sequence control goals, such as voltage balancing, as discussed in Sec.~\ref{sec:asym_faults}. 
In this light, in the second method, the negative-sequence reference is prioritized such that the available current headroom is first used to apply the negative-sequence component, and then the residual headroom (if available) is used for the positive-sequence component.
Thus, the negative-sequence reference is adjusted as
\begin{equation*}
	\B{\bar{i}}_{\dq-}^{\mathrm{ref}} = \begin{cases}
		\B{i}_{\dq-}^{\mathrm{ref}}, & I_\mathrm{max}\leq I_\mathrm{lim},\\
		\B{i}_{\dq-}^{\mathrm{ref}}, & I_\mathrm{max}\geq I_\mathrm{lim}, \quad||\B{i}_{\dq-}^{\mathrm{ref}}|| < I_\mathrm{lim}, \\
		I_\mathrm{lim}\frac{\B{i}_{\dq-}^{\mathrm{ref}}}{||\B{i}_{\dq-}^{\mathrm{ref}}||}, & I_\mathrm{max} \geq I_\mathrm{lim}, \quad ||\B{i}_{\dq-}^{\mathrm{ref}}||\geq I_\mathrm{lim},
	\end{cases}
\end{equation*}
with $I_\mathrm{max}=\max\{I_{\mathrm{max},\mathrm{a}},I_{\mathrm{max},\mathrm{b}},I_{\mathrm{max},\mathrm{c}}\}$, whereas the positive-sequence current reference is then calculated by
\begin{equation}
	\B{\bar{i}}_{\dq+}^{\mathrm{ref}} = \begin{cases}
		\B{i}_{\dq+}^{\mathrm{ref}}, & I_\mathrm{max}\leq I_\mathrm{lim},\\
		\gamma'\B{i}_{\dq+}^{\mathrm{ref}}, & I_\mathrm{max}\geq I_\mathrm{lim}, \quad||\B{i}_{\dq-}^{\mathrm{ref}}|| < I_\mathrm{lim}, \\
		0, & I_\mathrm{max} \geq I_\mathrm{lim}, \quad ||\B{i}_{\dq-}^{\mathrm{ref}}||\geq I_\mathrm{lim},
	\end{cases}
\end{equation}
where the scaling factor is computed as
\begin{equation*}
    \gamma' = \frac{\sqrt{\norm{\B{i}_{\dq+}^{\mathrm{ref}}}^2{(I_\mathrm{lim}^2-\norm{\B{i}_{\dq-}^{\mathrm{ref}}}^2)+R_{\hat{x}}^2}} -R_{\hat{x}}  }{\norm{\B{i}_{\dq+}^{\mathrm{ref}}}^2}.
\end{equation*}
The phase with maximum current amplitude is determined by $\hat{x}=\argmax_x\{I_{\mathrm{max},x}\},$ and then $R_{\hat{x}}=\mathrm{Re}\{\B{i}_{\dq+}^{\mathrm{ref}}\B{i}_{\dq-}^{\mathrm{ref}}e^{j2\lambda_{\hat{x}}} \}$.

\subsection{Current Control}\label{sec:current_control}

The control scheme described up to now is compatible with various current controllers, see~\cite[\S 12]{teodorescu_grid_2011} and the options therein. To realize the current references established in the previous section, we employ a stationary reference frame proportional current controller with a PCC voltage feedforward which also accounts for the voltage drop across the filtering resistance and inductance $R_\mathrm{f}$ and $L_\mathrm{f}$. It is given by:
\begin{equation}\label{eq:stationary_cc}
	\begin{split}
    \B{v}_\mathrm{ref} =\,&K_{\mathrm{cc,p}}(\B{R}(-\omega_\mathrm{r} t)\B{\bar{i}}_{\dq+}^{\mathrm{ref}}+\B{R}(\omega_\mathrm{r} t)\B{\bar{i}}_{\dq-}^{\mathrm{ref}}-\B{{i}}_\mathrm{c})\\
    &+ R_\mathrm{f}(\B{R}(-\omega_\mathrm{r} t)\B{\bar{i}}_{\dq+}^{\mathrm{ref}}+\B{R}(\omega_\mathrm{r} t)\B{\bar{i}}_{\dq-}^{\mathrm{ref}})\\
     &+ j\omega_\mathrm{r}L_\mathrm{f}(\B{R}(-\omega_\mathrm{r} t)\B{\bar{i}}_{\dq+}^{\mathrm{ref}}-\B{R}(\omega_\mathrm{r} t)\B{\bar{i}}_{\dq-}^{\mathrm{ref}})\\
    &+G_\mathrm{bp}^\mathrm{ff}(s)\B{v}_\mathrm{pcc}, 
	\end{split}
\end{equation}
where $K_{\mathrm{cc,p}}$ is the proportional gain, $\B{v}_{\mathrm{ref}}$ is the converter reference voltage (e.g., to be created via PWM), and
\begin{equation*}
   G_\mathrm{bp}^\mathrm{ff}(s) = \frac{\alpha_\mathrm{ff} s}{s^2 + \alpha_\mathrm{ff} s + \omega_\mathrm{r}^2},
\end{equation*}
is a band-pass filter with a bandwidth $\alpha_\mathrm{ff}$ centered around~$\pm\omega_\mathrm{r}$, again using an adaptive gain selection to account for the variations in $\omega_\mathrm{r}$.

When using an output filter, such as an LCL filter, damping loops can be integrated into the current control design to mitigate filter resonance. However, for the case studies under consideration, this is not necessary. Additionally, a resonator for integral action could be included to address model mismatch, although this is beyond the current scope. Finally, the voltage drop term might also be defined based on the current measurements; however, such an approach would require sequence separation of the current measurements.

\section{FRT Strategies under Symmetrical Faults} \label{sec:sym_faults}
This section has two main parts. Firstly, we outline grid-forming control objectives under symmetrical faults from the perspective of grid codes and system stability. We then propose three FRT strategies for GFVCC that can achieve the given control objectives and discuss their properties and implementation requirements. Leveraging the modularity of GFVCC, these solutions adapt the control structure to make it suitable for operation under faults. The first modification option entails reverting the control structure back to the basic vector current control architecture, the second option takes advantage of the fast response provided by the virtual synchronous condenser, whereas the third option introduces an additional controller to downregulate the back EMF voltage $V_\mathrm{v}$ and thus adapt the control scheme to fault conditions. The three considered options are illustrated in Table~\ref{tab:fault_modes}, together with a brief summary of their main properties and drawbacks.
\begin{table*}[!t]
	\centering
	\caption{Three GFVCC Modification Options with FRT Services Provision Capabilities.}
	\label{tab:fault_modes}
	\begin{threeparttable}
		\arrayrulecolor{black!10}
		\begin{tabular}{llll}
			\arrayrulecolor{black}
			\hline\hline
			\textbf{Modification Option} & {Vector Current Control}  & {Virtual Synchronous Condenser}  & {Voltage Downregulation} \\
			\hline
		\textbf{Illustration}	& \makecell[l]{\includegraphics[width=4.4cm]{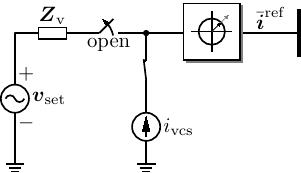}} & \makecell[l]{\includegraphics[width=4.4cm]{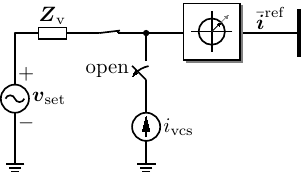}} & \makecell[l]{\includegraphics[width=4.4cm]{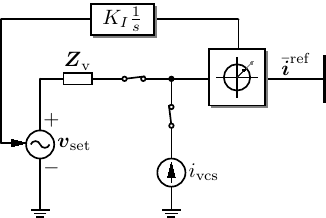}}\\
			\arrayrulecolor{black!10}
			\hline
			\textbf{Description} & \makecell[l]{Virtual sync. condenser is inactive, \\ and virtual current source is active.} & \makecell[l]{Virtual current source is inactive \\ and virtual sync. condenser is active.} & \makecell[l]{Additional integral controller $K_\mathrm{I}$ \\ downregulates back EMF voltage $V_\mathrm{v}$.} \\
			\hline
			\textbf{Synchronization} & {PLL bandwidth increased} & PLL freeze (and reset) & PLL preserved \\
			\hline
			\textbf{Voltage behavior} & $\norm{\B{v}_\mathrm{pcc}}$ following, $\angle{\B{v}_\mathrm{pcc}}$ following & $\norm{\B{v}_\mathrm{pcc}}$ following, $\angle{\B{v}_\mathrm{pcc}}$ forming & $\norm{\B{v}_\mathrm{pcc}}$ following, $\angle{\B{v}_\mathrm{pcc}}$ forming  \\
			\hline
            \textbf{Current behavior} & $\norm{\B{i}_\mathrm{c}}$ forming, $\angle{\B{i}_\mathrm{c}}$ forming & $\norm{\B{i}_\mathrm{c}}$ forming, $\angle{\B{i}_\mathrm{c}}$ following & $\norm{\B{i}_\mathrm{c}}$ forming, $\angle{\B{i}_\mathrm{c}}$ following  \\
			\hline
			\textbf{Max. fault current} & Yes, via setpoint adjustment & Yes, natural response & Yes, via integral control \\
			\hline
			 \textbf{Drawbacks} & Frequency runaway issues & \makecell[l]{Resynchronization is necessary \\ after fault since PLL is frozen.} & Controller tuning challenges \\
			\arrayrulecolor{black}
			\hline \hline
		\end{tabular}
        \begin{tablenotes}
        \item[1] The terminology for describing voltage and current behavior is adopted from \cite{Xiuqiang2024}.
        \end{tablenotes}
	\end{threeparttable}
    \vspace{-.25cm}
\end{table*}

\subsection{Grid-Forming Control Objectives}
The control objectives of grid-forming converters under grid faults have several layers. First of all, it is required that the converter stays connected to the grid and, ideally, provides all the grid-forming control functionalities as in normal operation. Nevertheless, as discussed earlier, grid-forming inverters must limit their output current, and thus, some functionalities might be obscured. Based on these observations, the control objectives addressed in this paper are outlined below.
\begin{enumerate}
    \item \textit{Control System Stability}: Once the fault occurs and the current limiter engages, it is common that the original converter setpoints can no longer be followed, implying a large discrepancy between the measured and the desired control signals. Therefore, a wind-up of integrators may occur and the control system stability may be jeopardized.  
    \item \textit{Fault Current Provision}: Considering that power system protection equipment uses fault current magnitude for fault detection, many grid codes worldwide require that the converter provides maximum allowable current (i.e., around 120\%-200\% rated current) during the entire fault duration. This requirement is typically achieved by maximizing the reactive power output of the converter during the fault while respecting the converter's current limits.
    \item \textit{Transient Stability}: Upon fault occurrence, an inverter should be able to regain an equilibrium operating point and remain synchronized with the rest of the system. Likewise, after fault clearing, it should be able to bring the voltage profile back to the nominal value and resynchronize with other units in the system.
\end{enumerate} 
It is important to note that the outlined control objectives define the scope of this paper; however, additional requirements may exist, such as ensuring certain harmonic power quality or addressing more specific objectives given in grid codes \cite{gb2024}.

\subsection{Fault Detection}
The fault detection is done via a hysteresis block that generates a flag signal once the voltage magnitude at the PCC falls below a predefined threshold, i.e., $\|\B{v}_{\mathrm{pcc},\mathrm{dq}+}\|\leq\underline{V}_\mathrm{th}$, and sets the fault flag back to zero once the voltage recovers $\|\B{v}_{\mathrm{pcc},\mathrm{dq}+}\|\geq\bar{V}_\mathrm{th}$. The thresholds need to be carefully selected as the voltage dip might be small in cases when the fault impedance is large. Furthermore, the thresholds directly impact delays in the control response at the fault occurrence and clearance. As a rule of thumb, one can select the trigger threshold $\underline{V}_\mathrm{th}$ to $80\%$ of the nominal voltage and recovery threshold $\bar{V}_\mathrm{th}$ to $75\%$ of the nominal system voltage. The generated fault signal is used to trigger the appropriate modifications in the control scheme (as discussed in the following) once the fault occurs, but also to revert the control scheme to its original form once the fault is cleared.

\subsection{Vector Current Control}
 The GFVCC concept was originally designed as an add-on to the standard vector current control \cite{VCC1998} to enhance its capabilities to provide grid supporting control features. The basic vector current control scheme possesses several desirable properties that enable it to operate under fault conditions: (i) current limiting is inherent, (ii) it is straightforward to impress maximum allowable (fault) current, and (iii) there is no danger of integrator wind-up due its grid-following behavior.

 \subsubsection{Control Modifications}
 Having in mind these benefits, one may consider reverting the GFVCC structure back to the basic vector current control structure during faults. This transformation of the control structure can be achieved by making the following adjustments once the fault is detected and indicated by the fault flag. Firstly, disregarding the virtual synchronous condenser reference $\B{i}_\mathrm{vsc,dq+}\rightarrow0$, implying that $\B{i}^\mathrm{ref}_{\dq+}=\B{i}_{\mathrm{vcs},\dq+}=0+jI_\mathrm{lim}$. The q-component of the virtual current source is set to $I_\mathrm{lim}$ to impress maximum allowable fault current. Secondly, the bandwidth of the PLL can optionally be increased by appropriately adjusting $K_{\mathrm{pll},\mathrm{p}},K_{\mathrm{pll},\mathrm{i}}$.
\subsubsection{Transient Stability Properties}
This solution, although simple to implement, suffers a drawback that is common to all grid-following schemes -- in fault scenarios when the grid is split by the fault such that no voltage sources remain in the part of the network with the converter, such as container faults, frequency runaway occurs since PLL has no voltage source to synchronize to \cite{SyncReview2020}. Approaches such as PLL freezing or keeping the PLL bandwidth low, as given by \eqref{eq:PLL_gains}, can be employed to alleviate this issue. The nominal PLL tuning of GFVCC results in a lower proportional gain
than that of typical PLLs found in grid-following controllers.

\subsection{Virtual Synchronous Condenser}\label{sec:vsc_approach}
The previous method for handling faults with GFVCC leverages its base vector current control structure; however, it does not maintain grid-forming properties during the fault and upon fault clearing. If the complete GFVCC structure is preserved during the fault instead, the following stability-related issues may arise: (i) wind-up of the AVR integrator in \eqref{eq:AVR} and (ii) large PLL frequency deviation. The first issue arises due to a potentially large difference between the measured PCC voltage magnitude $\norm{\B{v}_{\mathrm{pcc},\dq+}}$ and the setpoint magnitude  $V_\mathrm{v}$. The second issue arises from an instantaneous misalignment between the PLL-driven reference frame and the PCC voltage, triggered by the sudden change in the characteristics of the system. These two issues may develop during the fault and then cause resynchronization issues once the fault clears.
\subsubsection{Control Modifications}
Two simple control structure changes can be made to overcome the above-mentioned issues -- integrators within AVR in \eqref{eq:AVR} and PLL \eqref{eq:PLL} can be frozen and/or reset.
These changes have the following implications for the control scheme. By resetting and freezing the integrator within AVR, \eqref{eq:AVR} becomes $i_\mathrm{vsc,q+}=K_\mathrm{V}\cdot0=0$. Similarly, since the PLL integrator is frozen at a pre-fault condition,  we have that $\omega_\mathrm{r}=\omega_\mathrm{n}$, implying that the governor \eqref{eq:governor} becomes
\begin{equation}
    i_\mathrm{v,d+} = \frac{P_\mathrm{set}}{G_\mathrm{lpf}^\mathrm{v}(s)v_\mathrm{pcc,d+}}.
\end{equation}
Considering that active power supply is typically not required during the fault, an additional modification can be introduced, $P_\mathrm{set}=0$, which leads to $i_\mathrm{v,d+}=0$. Thus, the entire virtual current source is excluded from the control scheme $\B{i}_\mathrm{vcs}=0$. This is in contrast to the previous fault modification option, where the virtual current source provides the fault response, whereas the virtual synchronous condenser is removed. Note that although the virtual current response is zero, the maximum fault current is provided due to the large current flow through the virtual admittance, as discussed in the following. 
\subsubsection{Transient Stability Properties}
Given that the virtual admittance magnitude is typically $\norm{\B{Y}_\mathrm{v}}\gg1$ and that during most short-circuit faults $\norm{\B{v}_\mathrm{pcc,dq+}}\ll V_\mathrm{v}$, it follows from \eqref{eq:virtual_admittance} that $\norm{(V_\mathrm{v}\angle{0}-\B{v}_\mathrm{pcc,dq+})\B{Y}_\mathrm{v}}>I_\mathrm{lim}\approx1.2$. Under these conditions, the current limiter \eqref{eq:balanced_limiter} engages, and the impressed current vector adopts the following form:
\begin{equation}
    \bar{i}_{\dq+}^\mathrm{ref} = I_\mathrm{lim}\angle{\big((V_\mathrm{v}\angle{0}-\B{v}_\mathrm{pcc,dq+})\B{Y}_\mathrm{v}\big)},
\end{equation}
implying that the virtual synchronous condenser under faults behaves as a current source with current magnitude at $I_\mathrm{lim}$ and angle depending solely on the angle of $\B{v}_\mathrm{pcc,dq+}$ and on the virtual admittance selection. Therefore, the current magnitude is formed while the angle is a function of $\angle{\B{v}_\mathrm{pcc,dq+}}$. Given the current source behavior, there is no risk of transient instability concerning maximum power transfer limits and associated power angle curves \cite{Huang2019}. Additionally, upon fault clearance, the virtual synchronous condenser promptly restores the voltage to its nominal value as $\norm{V_\mathrm{v}\angle{0}-\B{v}_\mathrm{pcc,dq+}}\rightarrow0$.
 
\subsection{Voltage Downregulation}
A third modification option can be devised that retains both the virtual current source and virtual synchronous condenser functionalities, albeit at the cost of introducing an additional controller. Namely, the two issues that arise under faults are both related to the discrepancy between the PCC voltage magnitude and the setpoint magnitude $V_\mathrm{v}$. Instead of freezing the relevant integrators, the voltage setpoint $V_\mathrm{v}$ can be downregulated to minimize this discrepancy and avoid integrator wind-up, as discussed below.

\subsubsection{Control Modifications}
The objective is to construct an appropriate virtual back EMF voltage $V_\mathrm{v}$ by enforcing the current reference magnitude at the allowable limit, which can be achieved using an integral controller of the form: 
\begin{equation}
    V_\mathrm{v} = V_\mathrm{v0} - K_\mathrm{I}\frac{1}{s}\max(\norm{\B{i}_\mathrm{dq+}^\mathrm{ref}}-I_\mathrm{lim},0),
\end{equation}
where $K_\mathrm{I}$ is an integral regulator and $V_\mathrm{v0}$ is the initial voltage magnitude setpoint. The max-function is used in order to keep the voltage downregulation inactive in normal operation, i.e., before the current limit is reached. The angle reference $\theta_\mathrm{r}$ is generated by the PLL, as during normal operation, in order to construct the setpoint voltage vector $\B{v}_{\mathrm{set},\dq+} = V_\mathrm{v}\angle{\theta_\mathrm{r}}$. The remainder of the control scheme remains unchanged. Given that the voltage is adapted, there is no need to freeze or reset the integrators.
\subsubsection{Transient Stability Properties} Considering the aforementioned modifications, the angle-forming characteristic (synchronization properties) of GFVCC is preserved, and the voltage magnitude is utilized to impose the maximum allowable fault current provision. Consequently, the AC side behavior can be described as current magnitude and voltage angle forming. Transient stability can be analyzed formally using equivalent normal forms and extended transient analysis, as presented in \cite{Xiuqiang2024}, which examined a similar voltage downregulation scheme.

\section{FRT Strategies under Asymmetrical Faults} \label{sec:asym_faults}
In this section, we first outline the typical control goals for the negative-sequence stated in the grid codes or common in engineering practice. Subsequently, we describe several controller implementations that can achieve the control goals.  

\subsection{Negative-Sequence Control Objectives}
The FRT services in the negative-sequence typically admit multiple options, depending on grid codes or application-based specific requirements. The considered control goals can be classified into three groups:
\begin{enumerate}
    \item \textit{Converter Current Balancing}: Asymmetrical faults are always followed by negative-sequence current flows, implying that the injected currents of the converter are unbalanced if no additional negative-sequence control loops are implemented. To this end, in certain applications, it might be desirable to balance the converter currents.
    \item \textit{Power Oscillation Suppression}: When the output current and voltage waveforms include both positive- and negative-sequence components, active and reactive power exhibit oscillations at twice the fundamental frequency. These active power oscillations induce a ripple in the DC-link voltage, which may be unacceptable in certain applications. Consequently, negative-sequence controls are often designed to suppress these power oscillations.
    \item \textit{Negative-Sequence Voltage Mitigation}: Grid codes often mandate the injection of negative-sequence current to mitigate negative-sequence voltage magnitude and thereby reduce the voltage unbalance factor. Notably, IEEE Std. 2800-2022 \cite{ieee2800} specifies that all inverter-based resources must absorb negative-sequence reactive current in proportion to the negative-sequence voltage. This control objective has been extensively implemented in grid-following converters, as documented in \cite{Jundi2018}.
\end{enumerate}

\subsection{Negative-Sequence Controller Variants}
In this section, we discuss solutions that achieve the above control goals and instantiate the negative-sequence control block first introduced in Fig.~\ref{fig:gfvcc_overview}. It is furthermore important to note that, independent of which of the negative-sequence control strategies are implemented, one of the strategies discussed for symmetrical faults, i.e., vector current control, virtual synchronous condenser, or voltage downregulation, needs to be implemented in order to avoid the previously discussed control system stability issues. Therefore, a complete FRT strategy for GFVCC can be obtained by selecting one strategy from Sec.~\ref{sec:sym_faults} and one of the strategies discussed in the following.

\subsubsection{Balanced Current Control}
The output current of the converter is balanced when the negative-sequence current component is absent. This can simply be achieved by specifying the negative-sequence current reference to zero, as $\B{i}_{\dq-}^\mathrm{ref} = \B{0}$.

\subsubsection{Power Oscillation Suppression Control}
The oscillation in active power can be eliminated by appropriately specifying the negative-sequence current reference. More precisely, it was shown in \cite{Zheng2018} that the power oscillation is removed if and only if the current reference is given by:
\begin{equation}
	\B{i}_{\dq-}^\mathrm{ref} = -\frac{\B{v}_{\mathrm{pcc},\dq-}}{\mathrm{conj}(\B{v}_{\mathrm{pcc},\dq+})}\mathrm{conj}(\B{i}_{\dq+}^\mathrm{ref}),
\end{equation} 
where $\mathrm{conj}(\cdot)$ denotes the conjugation operation. Note that the negative-sequence current reference calculation relies on the positive-sequence current reference.

\subsubsection{Converter Voltage Balancing}
The negative-sequence voltage at the converter output (terminal) can be mitigated by introducing a virtual admittance in the negative-sequence circuit, described by
\begin{equation}
	\B{i}_{\dq-}^\mathrm{ref}=\B{Y}_\mathrm{v-}(\B{v}_{\mathrm{set},\dq-}-\B{v}_{\mathrm{pcc},\mathrm{dq}-}),
\end{equation}
where the negative-sequence voltage setpoint is set to zero $\B{v}_{\mathrm{set},\dq-}={0}$ given that the control goal is to mitigate its development after a fault.  The virtual admittance is implemented as a static admittance to avoid higher-order dynamics in the negative-sequence. The value for the virtual admittance $\B{Y}_\mathrm{v-}$ is recommended to be selected to correspond to the admittance between the converter terminal and PCC, defined by $R_\mathrm{f}, L_\mathrm{f}$ in Fig.~\ref{fig:system-model}.  Given that this admittance is typically inductive, this strategy corresponds with the absorption of negative-sequence reactive current for negative-sequence voltage mitigation. Higher values might fail to achieve the control goal, whereas significantly lower values might lead to instability.

\subsubsection{Active Damping}
The virtual admittance in negative-sequence would typically have only a small resistive component corresponding to $R_\mathrm{f}$. Although a larger resistive component would contribute to improved damping in the negative-sequence response, in steady-state it would create an error in the form of a non-zero negative-sequence voltage. Instead, a virtual termination resistor can be implemented via a high-pass filter for improved damping characteristics without causing steady-state errors. Therefore, an additional current reference can be superimposed to $\B{i}_{\dq-}^\mathrm{ref}$, computed as
\begin{equation}
	\boldsymbol{i}_{\mathrm{ad},\dq-} = \frac{1}{R_\mathrm{ad-}} {G}_\mathrm{hp}(s)\B{v}_{\mathrm{pcc},\dq-},
\end{equation}
where a high-pass filter ${G}_\mathrm{hp}(s)$ is applied to the PCC voltage measurement and subsequently used in conjunction with a virtual damping resistor to compute the reference. As a rule of thumb, the same value for $R_\mathrm{ad-}$ as for the virtual resistor in the positive-sequence can be used, i.e., $R_\mathrm{ad-}=R_\mathrm{ad}$.

\section{Results}\label{sec:res}
\begin{figure}[!b]
    \centering
    \includegraphics[scale=0.7]{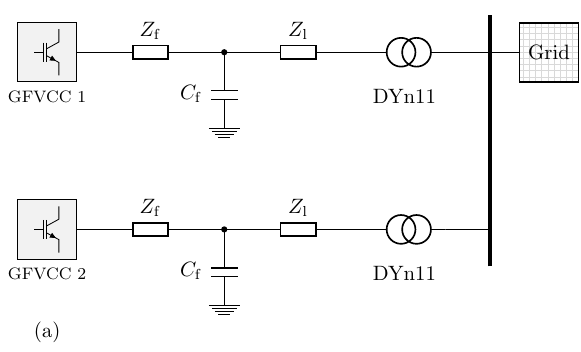}
    \includegraphics[scale=0.825]{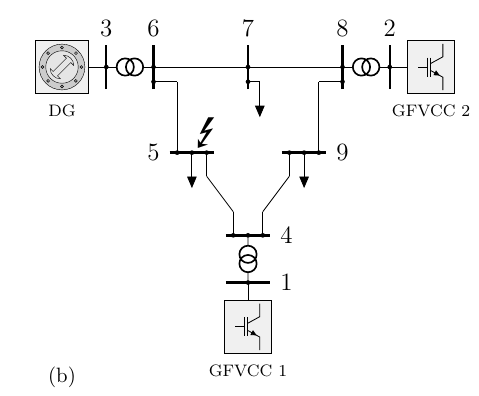}
    \caption{Single-line diagrams of the considered test systems: (a) Case Study I: Parallel inverters feeding a grid equivalent; (b) Case Study II: IEEE 9-bus system with two GFVCC converters and a diesel generator.}
    \label{fig:test_networks}
\end{figure}
In this section, the previously proposed symmetrical and asymmetrical FRT strategies are evaluated through two case studies, employing the test networks depicted in Fig.~\ref{fig:test_networks}. Case Study I examines a configuration in which two inverters operating in parallel are connected to an infinite bus. Case Study II considers a modified version of the IEEE 9-bus test system, with two generating units configured as converters operating in GFVCC mode and the third unit being a diesel generator (DG). Both test systems are implemented in \textsc{MATLAB} Simulink using blocks from the SimPowerSystems Library.

The converter system used in the case studies is composed of an LCL filter, a DYn11 transformer on the AC side, and a DC-link capacitor with a controlled current source that regulates the DC-link voltage on the DC side. The PCC (and the point where voltage is measured) is assumed to be the LCL filter capacitor node, as indicated in Fig.~\ref{fig:system-model}. While an averaged switching model is used, the switching and control system delays are taken into account. The converter control system employs the extended GFVCC scheme presented in Sec.~\ref{sec:GFVCC_overview}. The parameters of the physical converter components and the control system are summarized in Table~\ref{tab:parameters}. The employed PLL gains are designed to emulate a virtual inertia constant of $1$ second and a damping factor of $0.8$. The selected current control gain $K_\mathrm{cc,p}$ reflects its $700$\,Hz bandwidth.
\begin{table}[!t]
\renewcommand{\arraystretch}{1.1}
\centering
\caption{Converter and control system parameters}
\begin{tabular}[t]{lll}
\arrayrulecolor{black}\toprule
Description                 & Label       & Value               \\
\hline
\multicolumn{3}{c}{Converter Parameters} \\
\arrayrulecolor{black!30}\hline
Converter nominal power & $P_\mathrm{n}$ & $100$\,kW  \\
Nominal system voltage & $V_\mathrm{n}$ & $400$\,V  \\
Nominal angular frequency & $\omega_\mathrm{n}$ & $2\pi50$\,rad/s  \\
Filter impedance & $Z_\mathrm{f}$ & $0.002+j0.04$\,pu  \\
Line impedance & $Z_\mathrm{l}$ & $0.001+j0.02$\,pu  \\
Transformer impedance & $Z_\mathrm{t}$ & $0.002+j0.04$\,pu  \\
Switching frequency & - & $14$\,kHz  \\
\arrayrulecolor{black}\hline
\multicolumn{3}{c}{Control Parameters} \\
\arrayrulecolor{black!30}\hline
Virtual resistance pos.-sequence & $R_\mathrm{v}$ & $0.045$\,pu  \\
Virtual inductance pos.-sequence & $L_\mathrm{v}$ & $0.18$\,pu  \\
PLL proportional gain & $K_\mathrm{pll,p}$ & 0.1\,pu \\
PLL integral gain & $K_\mathrm{pll,i}$ & 1.4\,pu \\
Governor droop gain & $K_\mathrm{g}$ & $20$\,pu \\
AVR integral gain & $K_\mathrm{V}$ & $75$\,pu \\
Active damping pos.-sequence & $R_\mathrm{ad}$ & $0.66$\,pu   \\
Current control prop. gain & $K_\mathrm{cc,p}$ & $0.56$\,pu  \\
Feed-forward filter bandwidth & $\alpha_\mathrm{ff}$ & $200$\,Hz  \\
Current limit & $I_\mathrm{lim}$ & $1.2$\,pu \\
Downregulation integral gain & $K_\mathrm{I}$ & $10$ \,pu \\
Virtual admittance neg.-sequence & $Y_\mathrm{v-}$ & $-j0.04$\,pu  \\
\arrayrulecolor{black}\bottomrule
\end{tabular}
\label{tab:parameters}
\end{table}
\begin{figure*}[]
  \begin{center}
  \includegraphics[width=\textwidth]{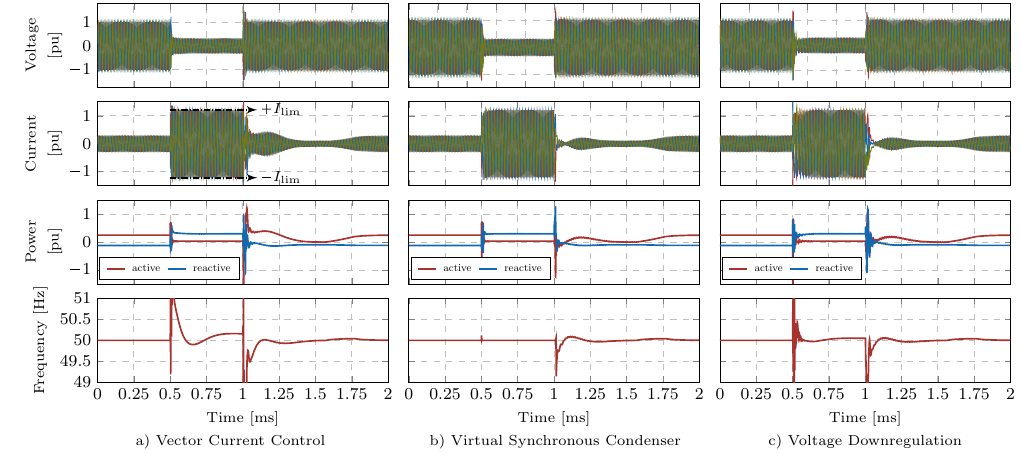}
  \caption{Case Study~I results for a symmetrical grid fault: a) vector current control FRT modification, b) virtual synchronous condenser FRT modification, c) voltage downregulation FRT strategy.}
  \label{fig:case-study-I}
  \end{center}
\end{figure*}

\subsection{Case Study I}
In this case study, we first analyze a symmetrical three-phase fault at the grid equivalent in the system of Fig.~\ref{fig:test_networks}a). The grid impedance is selected to reflect the short-circuit ratio $5$ and the XR ratio is $10$. 
The simulation results for GFVCC~1 for all three FRT strategies proposed in Sec.~\ref{sec:sym_faults} are presented in Fig.~\ref{fig:case-study-I}. Since GFVCC 1 and GFVCC 2 have the same configuration, their individual responses do not differ significantly. The simulation events are arranged in the following manner. At $t=0.5$\,s, a three-phase fault occurs at the infinite bus, the fault is cleared at $t=1$\,s, the power setpoint remains at zero for another $0.5$\,s to allow for resynchronization, and finally, a power ramp-up is initiated at $t=1.5$\,s. The active power setpoint is set at $0.25$\,pu to demonstrate that it can rapidly be increased to $I_\mathrm{lim}$ for maximum current provision during the fault. 

The results show that all three considered FRT strategies provide the desired fault current during the entire fault duration and that GFVCC remains stable and synchronized during and after the fault. Furthermore, all schemes provide active power post-fault to contribute to the grid energization. As can be seen from the plots, instantaneous overcurrent and overvoltage instances may occur at fault occurrence or fault clearance due to relatively small fault impedance and limited control bandwidth. Nevertheless, these instances do not pose significant risks for semiconductors and are addressed with switch-level limiters. Vector current control FRT modification exhibits the highest frequency oscillations since the control scheme has been reduced to PLL and current control only. Moreover, post-fault synchronization appears to be the slowest with this modification. On the other hand, the voltage downregulation scheme introduces additional dynamics due to the presence of the integrator, as can be seen in the voltage and power trajectories. The virtual synchronous condenser scheme delivers the smoothest response, making it the most favorable approach in this case study.
\begin{figure*}[]
  \begin{center}
  \includegraphics[width=\textwidth]{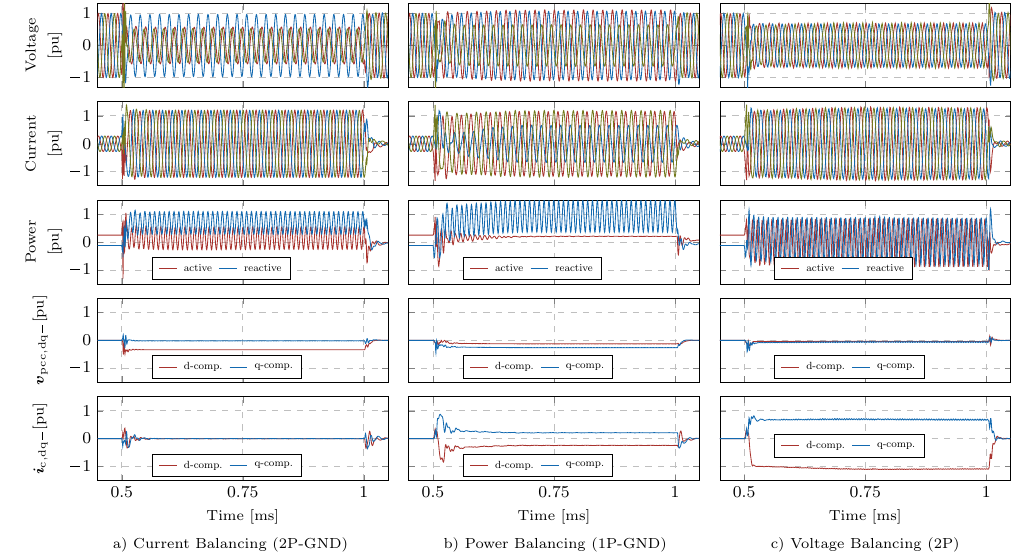}
  \caption{Case Study~I results for asymmetrical grid faults: a) current balancing control for the case of phase-to-phase-to-ground fault, b) power oscillation suppression control for the case of single phase-to-ground fault, c) voltage balancing control for the case of phase-to-phase fault.}
  \label{fig:case-study-I-asym}
  \end{center}
\end{figure*}
\begin{figure*}[]
  \begin{center}
  \includegraphics[width=\textwidth]{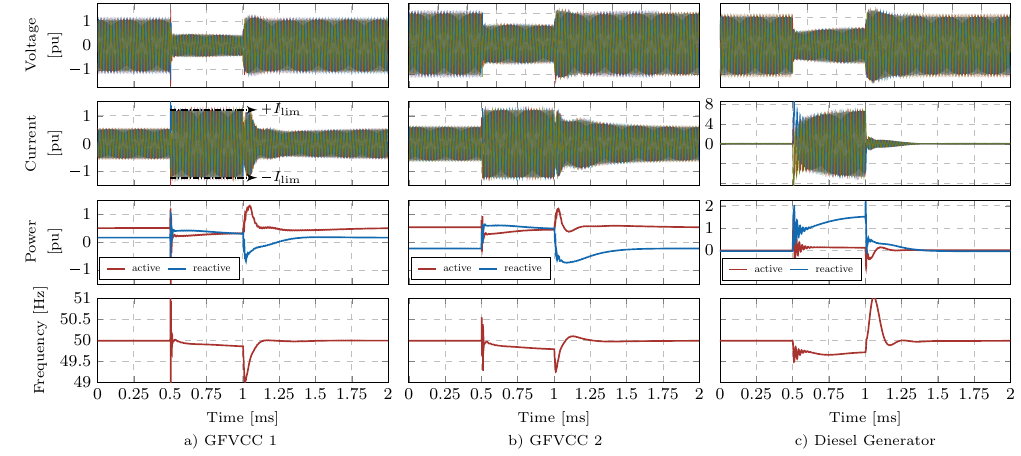}
  \caption{Case Study~II results, showing the trajectories of three-phase converter voltages, three-phase converter currents, active and reactive power injections, and frequency for a) GFVCC 1, b) GFVCC 2, and c) diesel generator.}
  \label{fig:case-study-II}
  \end{center}
\end{figure*}

Subsequently, we examine the performance of the proposed negative-sequence controllers under asymmetrical fault conditions, as introduced in Section \ref{sec:asym_faults}. The parameters of the case study remain consistent with those used in the symmetrical fault analysis. Based on the superior performance of the virtual synchronous condenser FRT strategy in the symmetrical fault cases, this strategy is applied to all asymmetrical fault simulations. To ensure that the specified control objectives are met, a negative-sequence priority current limiter is employed.

The system response for the three negative-sequence control strategies -- balanced current control, power oscillation suppression, and converter voltage balancing -- is presented in Fig.~\ref{fig:case-study-I-asym}. To cover a variety of asymmetrical faults, the different control strategies are simulated with different types of asymmetrical faults. The time axis is configured to emphasize the fault duration to focus on the period during which negative-sequence quantities are nonzero. The results demonstrate that the control objectives are achieved in all fault scenarios.

\subsection{Case Study II}
This case study aims to demonstrate the validity of the proposed FRT strategies in multi-unit networks without an infinite bus. The two GFVCC converters are configured identically to those in the previous case study, while the parameters and model of the DG are adopted from \cite{Schweizer2022}. The power setpoints of the two converters are set to $0.5$\,pu, whereas the DG operates with a zero power setpoint. The active power consumption of loads on buses 5, 7 and 9 is respectively $0.25$\,pu, $0.5$\,pu and $0.25$\,pu. The FRT strategy used for converters is the virtual synchronous condenser, with the PLL integrator being frozen instead of reset for an improved resynchronization performance. A symmetrical three-phase fault is initiated at bus 5 in Fig.~\ref{fig:test_networks} at $t=0.5$\,s for a duration of $0.5$\,s. The results, shown in Fig.~\ref{fig:case-study-II}, illustrate that immediately after the fault occurs, the converter current rapidly increases to $I_\mathrm{lim}$, while the DG current surges to values between $5$\,pu and $8$\,pu. During the fault, the DG exciter attempts to restore voltage, resulting in a gradual increase in its reactive power delivery and a slight rise in the active power output of the converters. After fault clearance, the system restores the previous steady state, governed by the DG's slower dynamic response.

\section{Conclusion \& Outlook} \label{sec:concl}
In this paper, we extended the grid-forming vector current control strategy to include negative-sequence control and proposed several fault ride-through (FRT) strategies for both symmetrical and asymmetrical faults. The results demonstrate that the proposed strategies effectively achieve FRT objectives, with the virtual synchronous condenser-based approach emerging as the most effective and robust solution. Conversely, reverting to the standard vector current control scheme was found to be the least favorable option. While the voltage downregulation strategy performed satisfactorily, it comes at the cost of increased complexity and commissioning effort. Nevertheless, the recommendation on the most suitable fault modification method for GFVCC depends also on the type of the network to which GFVCC is connected to. Namely, in weak grid conditions and islanded operation, the virtual synchronous condenser-based approach might be the most suitable approach, whereas in strong grids, reverting to the standard vector current control might be the simplest modification that provides the desired behavior.

A drawback of the proposed scheme is the necessity to modify the control structure and implement a fault detection mechanism. Consequently, the reliability of the proposed schemes is heavily contingent upon the dependability of the fault detection system employed. Ideally, a fault ride-through mode would be intrinsically integrated into the grid-forming method. Our future research will concentrate on developing such an approach.

There are several other important research directions that will be explored in our future work: (i) comparison between GFVCC and other commonly employed grid-forming methods under fault conditions, (ii) experimental validation of the FRT proposed approaches, and (iii) investigation of more reliable and accurate fault detection schemes.

% References section
\bibliographystyle{IEEEtran}
\bibliography{bibliography}

\begin{thebibliography}{10}
\providecommand{\url}[1]{#1}
\csname url@samestyle\endcsname
\providecommand{\newblock}{\relax}
\providecommand{\bibinfo}[2]{#2}
\providecommand{\BIBentrySTDinterwordspacing}{\spaceskip=0pt\relax}
\providecommand{\BIBentryALTinterwordstretchfactor}{4}
\providecommand{\BIBentryALTinterwordspacing}{\spaceskip=\fontdimen2\font plus
\BIBentryALTinterwordstretchfactor\fontdimen3\font minus \fontdimen4\font\relax}
\providecommand{\BIBforeignlanguage}[2]{{%
\expandafter\ifx\csname l@#1\endcsname\relax
\typeout{** WARNING: IEEEtran.bst: No hyphenation pattern has been}%
\typeout{** loaded for the language `#1'. Using the pattern for}%
\typeout{** the default language instead.}%
\else
\language=\csname l@#1\endcsname
\fi
#2}}
\providecommand{\BIBdecl}{\relax}
\BIBdecl

\bibitem{Yao2017}
Y.~Sun, X.~Hou, J.~Yang, H.~Han, M.~Su, and J.~M. Guerrero, ``New perspectives on droop control in ac microgrid,'' \emph{IEEE Transactions on Industrial Electronics}, vol.~64, no.~7, pp. 5741--5745, 2017.

\bibitem{PSC2010}
L.~Zhang, L.~Harnefors, and H.-P. Nee, ``Power-synchronization control of grid-connected voltage-source converters,'' \emph{IEEE Transactions on Power Systems}, vol.~25, no.~2, pp. 809--820, 2010.

\bibitem{VSM2015}
S.~D’Arco, J.~A. Suul, and O.~B. Fosso, ``A virtual synchronous machine implementation for distributed control of power converters in smart grids,'' \emph{Electric Power Systems Research}, vol. 122, pp. 180--197, 2015.

\bibitem{VOC2014}
B.~B. Johnson, S.~V. Dhople, A.~O. Hamadeh, and P.~T. Krein, ``Synchronization of parallel single-phase inverters with virtual oscillator control,'' \emph{IEEE Transactions on Power Electronics}, vol.~29, no.~11, pp. 6124--6138, 2014.

\bibitem{Schweizer2022}
M.~Schweizer, S.~Almér, S.~Pettersson, A.~Merkert, V.~Bergemann, and L.~Harnefors, ``Grid-forming vector current control,'' \emph{IEEE Transactions on Power Electronics}, vol.~37, no.~11, pp. 13\,091--13\,106, 2022.

\bibitem{VCC1998}
M.~Kazmierkowski and L.~Malesani, ``Current control techniques for three-phase voltage-source pwm converters: a survey,'' \emph{IEEE Transactions on Industrial Electronics}, vol.~45, no.~5, pp. 691--703, 1998.

\bibitem{Sadeghkhani2017}
I.~Sadeghkhani, M.~E. Hamedani~Golshan, J.~M. Guerrero, and A.~Mehrizi-Sani, ``A current limiting strategy to improve fault ride-through of inverter interfaced autonomous microgrids,'' \emph{IEEE Transactions on Smart Grid}, vol.~8, no.~5, pp. 2138--2148, 2017.

\bibitem{Moawwad2014}
A.~Moawwad, M.~S. El~Moursi, and W.~Xiao, ``A novel transient control strategy for vsc-hvdc connecting offshore wind power plant,'' \emph{IEEE Transactions on Sustainable Energy}, vol.~5, no.~4, pp. 1056--1069, 2014.

\bibitem{Du2023}
W.~Du and S.~M. Mohiuddin, ``A two-stage current limiting control strategy for improved low-voltage ride-through capability of direct-droop-controlled, grid-forming inverters,'' in \emph{2023 IEEE Energy Conversion Congress and Exposition (ECCE)}, 2023, pp. 2886--2890.

\bibitem{Paquette2015}
A.~D. Paquette and D.~M. Divan, ``Virtual impedance current limiting for inverters in microgrids with synchronous generators,'' \emph{IEEE Transactions on Industry Applications}, vol.~51, no.~2, pp. 1630--1638, 2015.

\bibitem{Liu2022}
T.~Liu, X.~Wang, F.~Liu, K.~Xin, and Y.~Liu, ``A current limiting method for single-loop voltage-magnitude controlled grid-forming converters during symmetrical faults,'' \emph{IEEE Transactions on Power Electronics}, vol.~37, no.~4, pp. 4751--4763, 2022.

\bibitem{Baeckeland2024}
N.~Baeckeland, D.~Chatterjee, M.~Lu, B.~Johnson, and G.-S. Seo, ``Overcurrent limiting in grid-forming inverters: A comprehensive review and discussion,'' \emph{IEEE Transactions on Power Electronics}, vol.~39, no.~11, pp. 14\,493--14\,517, 2024.

\bibitem{Huang2019}
L.~Huang, H.~Xin, Z.~Wang, L.~Zhang, K.~Wu, and J.~Hu, ``Transient stability analysis and control design of droop-controlled voltage source converters considering current limitation,'' \emph{IEEE Transactions on Smart Grid}, vol.~10, no.~1, pp. 578--591, 2019.

\bibitem{Lin2021}
X.~Lin, Y.~Zheng, Z.~Liang, and Y.~Kang, ``The suppression of voltage overshoot and oscillation during the fast recovery process from load short-circuit fault for three-phase stand-alone inverter,'' \emph{IEEE Journal of Emerging and Selected Topics in Power Electronics}, vol.~9, no.~1, pp. 858--871, 2021.

\bibitem{Rosso2021}
R.~Rosso, S.~Engelken, and M.~Liserre, ``On the implementation of an {FRT} strategy for grid-forming converters under symmetrical and asymmetrical grid faults,'' \emph{IEEE Transactions on Industry Applications}, vol.~57, no.~5, pp. 4385--4397, 2021.

\bibitem{ieee2800}
{IEEE Standards Association}, ``{IEEE} standard for interconnection and interoperability of inverter-based resources {(IBRs)} interconnecting with associated transmission electric power systems,'' \emph{IEEE Std. 2800-2022}, pp. 1--180, 2022.

\bibitem{NS-VSM2022}
E.~B. Avdiaj, S.~D’Arco, L.~Piegari, and J.~A. Suul, ``Negative sequence control for virtual synchronous machines under unbalanced conditions,'' \emph{IEEE Journal of Emerging and Selected Topics in Power Electronics}, vol.~10, no.~5, pp. 5670--5685, 2022.

\bibitem{DSVOC2023}
M.~A. Awal, M.~R.~K. Rachi, H.~Yu, I.~Husain, and S.~Lukic, ``Double synchronous unified virtual oscillator control for asymmetrical fault ride-through in grid-forming voltage source converters,'' \emph{IEEE Transactions on Power Electronics}, vol.~38, no.~6, pp. 6759--6763, 2023.

\bibitem{Bhagwat2023}
P.~Bhagwat and D.~Groß, ``Three-phase grid-forming droop control for unbalanced systems and fault ride through,'' in \emph{2023 IEEE Power \& Energy Society General Meeting (PESGM)}, 2023, pp. 1--5.

\bibitem{teodorescu_grid_2011}
R.~Teodorescu, M.~Liserre, and P.~Rodr{\'i}guez, \emph{\BIBforeignlanguage{eng}{Grid converters for photovoltaic and wind power systems}}, 1st~ed.\hskip 1em plus 0.5em minus 0.4em\relax Chichester: Wiley, 2011.

\bibitem{AwalSequences2022}
M.~A. Awal, M.~R.~K. Rachi, H.~Yu, S.~Schröder, and I.~Husain, ``Symmetrical components extraction for grid-forming voltage source converters,'' in \emph{2022 IEEE Energy Conversion Congress and Exposition (ECCE)}, 2022, pp. 1--8.

\bibitem{VSM2007}
H.-P. Beck and R.~Hesse, ``Virtual synchronous machine,'' in \emph{2007 9th International Conference on Electrical Power Quality and Utilisation}, 2007, pp. 1--6.

\bibitem{BoFan2022}
B.~Fan, T.~Liu, F.~Zhao, H.~Wu, and X.~Wang, ``A review of current-limiting control of grid-forming inverters under symmetrical disturbances,'' \emph{IEEE Open Journal of Power Electronics}, vol.~3, pp. 955--969, 2022.

\bibitem{Xiuqiang2024}
X.~He, M.~A. Desai, L.~Huang, and F.~Dörfler, ``Cross-forming control and fault current limiting for grid-forming inverters,'' \emph{IEEE Transactions on Power Electronics}, vol.~40, no.~3, pp. 3980--4007, 2025.

\bibitem{gb2024}
{National Grid ESO}, ``The grid code,'' National Grid Electricity System Operator, Tech. Rep., 2024.

\bibitem{SyncReview2020}
X.~Wang, M.~G. Taul, H.~Wu, Y.~Liao, F.~Blaabjerg, and L.~Harnefors, ``Grid-synchronization stability of converter-based resources—an overview,'' \emph{IEEE Open Journal of Industry Applications}, vol.~1, pp. 115--134, 2020.

\bibitem{Jundi2018}
J.~Jia, G.~Yang, and A.~H. Nielsen, ``A review on grid-connected converter control for short-circuit power provision under grid unbalanced faults,'' \emph{IEEE Transactions on Power Delivery}, vol.~33, no.~2, pp. 649--661, 2018.

\bibitem{Zheng2018}
T.~Zheng, L.~Chen, Y.~Guo, and S.~Mei, ``Flexible unbalanced control with peak current limitation for virtual synchronous generator under voltage sags,'' \emph{Journal of Modern Power Systems and Clean Energy}, vol.~6, no.~1, pp. 61--72, 2018.

\end{thebibliography}

% That's all folks
\end{document}